\documentclass[english,showpacs,english,aps,prc,twocolumn,superscriptaddress,floatfix]{revtex4}
\usepackage[T1]{fontenc}
\usepackage[latin9]{inputenc}
\usepackage{refstyle}
\usepackage{textcomp}
\usepackage{amstext}
\usepackage{amssymb}
\usepackage{graphicx}
\usepackage{color}
\usepackage{lipsum}
\usepackage{float} 
\usepackage{tabularx}
\usepackage{mathtools}
\usepackage{ulem}		

\makeatletter

\newcommand{\lyxmathsym}[1]{\ifmmode\begingroup\def\b@ld{bold}
  \text{\ifx\math@version\b@ld\bfseries\fi#1}\endgroup\else#1\fi}

\RS@ifundefined{subref}
  {\def\RSsubtxt{section~}\newref{sub}{name = \RSsubtxt}}
  {}
\RS@ifundefined{thmref}
  {\def\RSthmtxt{theorem~}\newref{thm}{name = \RSthmtxt}}
  {}
\RS@ifundefined{lemref}
  {\def\RSlemtxt{lemma~}\newref{lem}{name = \RSlemtxt}}
  {}

\@ifundefined{textcolor}{}
{%
 \definecolor{BLACK}{gray}{0}
 \definecolor{WHITE}{gray}{1}
 \definecolor{RED}{rgb}{1,0,0}
 \definecolor{GREEN}{rgb}{0,1,0}
 \definecolor{BLUE}{rgb}{0,0,1}
 \definecolor{CYAN}{cmyk}{1,0,0,0}
 \definecolor{MAGENTA}{cmyk}{0,1,0,0}
 \definecolor{YELLOW}{cmyk}{0,0,1,0}
}

\makeatother

\usepackage{babel}
\begin{document}

\title{Experimental study of the isospin transport with $^{40,48}$Ca+$^{40,48}$Ca reactions at 35 MeV/nucleon}

\author{Q.~Fable}
\email{quentin.fable@l2it.in2p3.fr}
\affiliation{Laboratoire des 2 Infinis - Toulouse (L2IT-IN2P3), Universit\'e de Toulouse, CNRS, UPS, F-31062 Toulouse Cedex 9 (France)}

\author{A. Chbihi}
\affiliation{GANIL, CEA/DRF-CNRS/IN2P3, Bvd. Henri Becquerel, F-14076 Caen CEDEX, France}

\author{J.D.~Frankland}
\affiliation{GANIL, CEA/DRF-CNRS/IN2P3, Bvd. Henri Becquerel, F-14076 Caen CEDEX, France}

\author{P.~Napolitani}
\affiliation{Universit\'e Paris-Saclay, CNRS/IN2P3, IJCLab, 91405 Orsay, France}

\author{G. Verde}
\affiliation{Laboratoire des 2 Infinis - Toulouse (L2IT-IN2P3), Universit\'e de Toulouse, CNRS, UPS, F-31062 Toulouse Cedex 9 (France)}
\affiliation{Istituto Nazionale di Fisica Nucleare, Sezione di Catania, 64 Via Santa Sofia, I-95123, Catania, Italy}


\author{E. Bonnet}
\affiliation{SUBATECH UMR 6457, IMT Atlantique, Universit\'e de Nantes, CNRS-IN2P3, 44300 Nantes, France}

\author{B.~Borderie}
\affiliation{Universit\'e Paris-Saclay, CNRS/IN2P3, IJCLab, 91405 Orsay, France}

\author{R.~Bougault}
\affiliation{Normandie Univ, ENSICAEN, UNICAEN, CNRS/IN2P3, LPC Caen, F-14000 Caen, France}

\author{E.~Galichet}
\affiliation{Universit\'e Paris-Saclay, CNRS/IN2P3, IJCLab, 91405 Orsay, France}
\affiliation{Conservatoire National des Arts et M\'etiers, F-75141 Paris Cedex 03, France}

\author{T.~G\'enard}
\affiliation{GANIL, CEA/DRF-CNRS/IN2P3, Bvd. Henri Becquerel, F-14076 Caen CEDEX, France}

\author{D.~Gruyer}
\affiliation{Normandie Univ, ENSICAEN, UNICAEN, CNRS/IN2P3, LPC Caen, F-14000 Caen, France}

\author{M.~Henri}
\affiliation{GANIL, CEA/DRF-CNRS/IN2P3, Bvd. Henri Becquerel, F-14076 Caen CEDEX, France}

\author{M.~La~Commara}
\affiliation{Dipartimento di Farmacia, Universit\`{a} Federico II and INFN Napoli, Napoli, Italia}

\author{A.~Le~F\`evre}
\affiliation{GSI Helmholtzzentrum f\"{u}r Schwerionenforschung GmbH, D-64291 Darmstadt, Germany}

\author{J.~Lemari\'e}
\affiliation{GANIL, CEA/DRF-CNRS/IN2P3, Bvd. Henri Becquerel, F-14076 Caen CEDEX, France}

\author{N.~Le~Neindre}
\affiliation{Normandie Univ, ENSICAEN, UNICAEN, CNRS/IN2P3, LPC Caen, F-14000 Caen, France}

\author{O.~Lopez}
\affiliation{Normandie Univ, ENSICAEN, UNICAEN, CNRS/IN2P3, LPC Caen, F-14000 Caen, France}

\author{M.~P\^arlog}
\affiliation{Normandie Univ, ENSICAEN, UNICAEN, CNRS/IN2P3, LPC Caen, F-14000 Caen, France}
\affiliation{National Institute for Physics and Nuclear Engineering, RO-077125 Bucharest-M\u{a}gurele, Romania}

\author{A.~Rebillard-Souli\'e}
\affiliation{Normandie Univ, ENSICAEN, UNICAEN, CNRS/IN2P3, LPC Caen, F-14000 Caen, France}

\author{E. Vient}
\affiliation{Normandie Univ, ENSICAEN, UNICAEN, CNRS/IN2P3, LPC Caen, F-14000 Caen, France}

\author{M.~Vigilante}
\affiliation{Dipartimento di Fisica, Universit\`a degli Studi di Napoli FEDERICO II, I-80126 Napoli, Italy}
\affiliation{Istituto Nazionale di Fisica Nucleare, Sezione di Napoli, Complesso Universitario di Monte S. Angelo, Via Cintia Edificio 6, I-80126 Napoli, Italy}

\collaboration{INDRA collaboration}\noaffiliation

\begin{abstract}
We investigate the isospin transport with $^{40,48}$Ca$+^{40,48}$Ca reactions at $35$ MeV/nucleon, measured with the coupling of the VAMOS high acceptance spectrometer and the INDRA charged particle multidetector.
Using the quasi-projectile remnant measured with VAMOS and carefully selected light-charged-particles measured in coincidence with INDRA, a reconstruction method is applied to estimate the excited quasi-projectile (QP) on an event-by-event basis.
The isospin diffusion is investigated using the isospin transport ratio with the asymmetry $\delta=(N-Z)/A$ of the projectile as an isospin-sensitive observable and the total transverse energy of $Z \leq 2$ nuclei for experimental centrality sorting.
The experimental isospin transport ratios present, for both the reconstructed QP and its remnant, a clear evolution towards isospin equilibration with increasing dissipation of the collision while the full equilibration is not reached. 
A smoother evolution with less discrepancies between the two mixed systems in the case of the reconstructed QP is also observed. 
The isospin migration is investigated using the neutron-to-proton ratio of the light-charged-clusters and the velocity of the QP remnant as a sorting parameter. 
More particularly, we focused on an angular region centered around the mid-rapidity of the reaction so as to characterize the low-density neck emissions.
A systematic neutron-enrichment is observed and interpreted as a consequence of isospin migration, more particularly for the symmetric systems which present no isospin gradient between the projectile and the target.
We also noticed that the $^{2}$H and $^{4}$He particles exhibit very close multiplicities independently of the sorting variable for the symmetric systems.
\end{abstract}
\date{\today}
\maketitle

\section{Introduction}\label{sec_intro}

The equation of state (NEoS) of bulk nuclear matter over a wide range of densities, temperatures and neutron-to-proton asymmetries remains a major issue in modern nuclear physics and astrophysics.
Indeed, the NEoS plays a key role in describing nuclear systems probed in laboratory experiments, as well as in modeling neutron stars, core-collapse supernovae (CCSN) and mergers of compact binary stars \cite{LATTIMER2016127}.
Heavy-ion collisions (HIC) are the unique tool to probe the NEoS at finite temperature under laboratory controlled conditions, over a wide range of density and energy, depending on the beam energy, the size of the colliding systems and the impact parameter of the collisions.
Such experiments allow to constrain the isoscalar (same proton, $\rho_p$, and neutron, $\rho_n$, densities) and isovector ($\rho_p \neq \rho_n$) contributions to the energy density of nuclear matter away from saturation density, $\rho_0 \simeq 0.17$ fm \cite{Tsang2009:ConstraintdensitydependenceSymm}.
At intermediate beam energies ($20-100$ MeV/nucleon), temperatures and densities similar to those reached in the neutrinosphere of CCSN matter are expected, allowing the experimental study of in-medium effects of light clusters at high temperature \cite{Pais_PRC_99_055806}.
At higher beam energies, NEoS constraints for symmetric nuclear matter \cite{LEFEVRE2016112} and symmetry energy \cite{Russotto_PhysRevC_94_034608} are probed at densities in the range of roughly $\rho_0\text{-to-}3\rho_0$ and $\rho_0\text{-to-}2\rho_0$, respectively. 
In particular, a recent interdisciplinary analysis from Huth \textit{et.~al} demonstrated the interest of combining astrophysical multi-messenger observations from gravitational-wave astronomy \cite{Abbott_PhysRevLett_119_161101}, electromagnetic observations of neutron stars \cite{Miller_2019} and HIC results to provide complementary information on the NEoS at intermediate densities \cite{Huth2022}.

As the NEoS cannot be measured directly, it can only be inferred from comparisons of model predictions with carefully selected collision observables.
For example, transport models are used to predict the dynamics of HIC by numerical resolution of equations based on semi-classical mean-field approximations, including correlations via nucleon-nucleon collisions and a given parametrization of the NEoS via effective interactions.
Peripheral and semi-peripheral reactions around Fermi energy ($30-60$ MeV/nucleon) are of particular interest to study the transport of nuclear matter and $N/Z$ equilibration effects.
Indeed such reactions are dominated by binary-like collisions, where projectile and target nuclei interact by exchanging nucleons before re-separating into a quasiprojectile (QP) and a quasitarget (QT), with kinematic properties respectively close to the projectile and the target, but also by the production of lighter ejectiles like light-charged particles (LCPs, $Z=1,2$) and intermediate-mass fragments (IMFs, $3 \leq Z \leq 6$ in the present work) \cite{BARAN2004329, PhysRevC_72_064620, PhysRevC_79_064615}. 
As QP and QT can be moderately deformed and excited, they are expected to undergo sequential decays by emitting LCPs and gamma rays, leading to a projectile-like (PLF) and a target-like (TLF) fragment as remnants.

If the reaction partners present a different initial neutron-to-proton ratio $N/Z$, they are expected to undergo isospin equilibration which will be reflected in the $N/Z$ of the outgoing nuclei.
This phenomenon, called isospin diffusion, is driven by the isospin gradient between the two colliding participants, which is proportional to the symmetry energy term of the NEoS.   
In addition, various transport model calculations suggest a neutron enrichment of the neck region developing between the QP and the QT and characterized by sub-saturation densities ($\rho < \rho_0$) \cite{Tsang2004:isospindiffusion, CHEN_PLR_94_032701, BARAN2005335, LIONTI200533, Tsang2009:ConstraintdensitydependenceSymm, Coupland:InfluenceTransportVariablesIsospin, Colonna_EPJA50}.
This phenomenon, called isospin migration (or drift), is driven by the density gradient between this neck region and the QP-QT remnants, which is proportional to the derivative of the symmetry energy term: a larger neutron enrichment of the neck is expected for a stiffer symmetry energy \cite{Colonna_EPJA50}.
One can thus expect the degree of charge and mass equilibration occuring during the collision mainly to depend on the strength of the symmetry energy and the interaction time between the two reaction participants.
Exploiting large angular acceptance detector arrays with different projectile and target combinations, it is possible to study the competition between isospin diffusion and migration while characterizing the dissipation of the collision. 
Of course, the time scales for each collision cannot be measured experimentally but only be inferred from measured final state observables.
To draw conclusions, it is also necessary to consider the role of evaporation and prompt-emitted particles that possibly modify the experimental observables.

This work presents the experimental investigation of the isospin transport in $^{48,40}$Ca$+^{48,40}$Ca reactions at $35$ MeV/nucleon measured with the INDRA-VAMOS coupling and follows our previous study reported in \cite{Fable_PRC_106_024605}.

The experimental setup is described in Section \ref{sec_ExpSetup}.
Section \ref{sec_IsospinTransport} presents the results on isospin diffusion and isospin migration, along with a comparison between the PLF measured in VAMOS and the reconstructed QP. 
Summary and conclusions are finally presented in Section \ref{sec_conclusion}.

\section{Experimental setup and QP reconstruction\label{sec_ExpSetup}}

The INDRA-VAMOS coupled setup was used at the GANIL facility to measure $^{48,40}$Ca$+^{48,40}$Ca collisions at 35 MeV/nucleon.

The VAMOS high acceptance spectrometer \cite{Pullanhiotan2008343:VAMOS} covered the forward polar angles from $2.56^{\circ}$ to $6.50^{\circ}$, so as to detect a single fragment emitted slightly above the grazing angle of the reaction. 
Focal plane detection capabilities consisted of a 7-modules ionization chamber, a 500 $\mu m$ thick Si-wall (18 independent modules) and a $1$ $cm$ thick CsI(Tl)-wall (80 independent modules), allowing the measurements of the time of flight, energy loss ($\Delta E$) and energy ($E$) parameters. 
Two position-sensitive drift chambers were used to determine the trajectories of the reaction products at the focal plane.
Around twelve magnetic rigidity ($B\rho_0$) settings, from $0.661$ to $2.220$ $T\,m$, were measured for each system to cover the full velocity range of the fragments and identify isotopically the forward-emitted PLF.

INDRA covered polar angles from $7^{\circ}$ to $176^{\circ}$ with detection telescopes arranged in rings centered around the beam axis. 
The forward rings ($7^{\circ}-45^{\circ}$) consisted each of three-layers detection telescopes: a gas-ionization chamber, a 300 or 150 $\mu m$ silicon wafer and a CsI(Tl) scintillator (14 to 10 $cm$ thick) read by a photomultiplier tube. 
The backward rings ($45^{\circ}-176^{\circ}$) included two-layers telescopes: a gas-ionization chamber and a CsI(Tl) scintillator (5 to 8 $cm$ thick). 
INDRA allowed charge and isotope identification up to Be-B and only charge identification for heavier fragments.

Further details of the setup, trajectory reconstruction, normalization procedure and QP reconstruction can be found in \cite{Fable_PRC_106_024605}, where the analysis allowed us to also estimate how to account for the undetected neutrons emitted by the QP.

\section{Isospin transport}\label{sec_IsospinTransport}

The stochastic, differential exchange of nucleons between two colliding nuclei with different $N/Z$, usually referred to as isospin transport, is expected to depend mainly on the difference in composition of the two nuclei, the interaction time of the collision, and the symmetry energy term of the NEoS.
In a hydrodynamic approximation, the isospin transport can be described by means of the local neutron and proton currents $\mathbf{j}_n$ and $\mathbf{j}_p$ , such as \cite{PhysRevC_72_064620, Colonna_EPJA50}:
\begin{equation}
\mathbf{j}_{n,p} = D^{\rho}_{n,p} \mathbf{\nabla}\rho - D^{\delta}_{n,p} \mathbf{\nabla}\delta
\label{eq_jnp} 
\end{equation}

where $\rho = \rho_n + \rho_p$ is the local density of nuclear matter, $\delta = \left( \rho_n-\rho_p \right)/\rho $ the neutron-to proton asymmetry (isospin) and $D^{\rho,\delta}_{n,p}$ are the density and isospin-dependent transport coefficients for neutrons and protons \cite{PhysRevC_72_064620}.

According to Eq.\ref{eq_jnp}:
\begin{equation}
\mathbf{j}_{n} - \mathbf{j}_{p} = \underbrace{\left( D^{\rho}_{n} - D^{\rho}_{p} \right) \mathbf{\nabla}\rho}_{\text{Isospin Migration}} - \underbrace{\left( D^{\delta}_{n} - D^{\delta}_{p} \right) \mathbf{\nabla}\delta}_{\text{Isospin Diffusion}} 
\label{eq_jn_m_jp} 
\end{equation}

The first term of Eq.\ref{eq_jn_m_jp}, called isospin migration (or isospin drift) can be related to the density dependence of the symmetry energy, as:

\begin{equation}
D^{\rho}_{n} - D^{\rho}_{p} \propto 4 \delta \frac{\partial \epsilon_{sym}}{\partial \rho}
\label{eq_csym_migration} 
\end{equation}

The second term of Eq.\ref{eq_jn_m_jp}, called isospin diffusion can be related to the value of the symmetry energy (at a given local density), as:

\begin{equation}
D^{\delta}_{n} - D^{\delta}_{p} \propto 4 \rho \epsilon_{sym}
\label{eq_csym_diffusion} 
\end{equation}

During a collision, a rearrangement of the neutron-to-proton ratio of the colliding nuclei is expected to take place under the combined effect of
(i) isospin migration (drift) leading to an increase in neutron richness of low-density regions and
(ii) isospin diffusion leading to a balance in neutron-richness of different isospin ($\delta$) regions.
Equations \ref{eq_csym_migration} and \ref{eq_csym_diffusion} hence suggest that one can use the measurements of $N/Z$-sensitive observables to probe the derivative and strength of $\epsilon_{sym}$, respectively. 

\subsection{Isospin diffusion}\label{subsec_IsospinDiffusion}

Quantitative signals of the isospin diffusion phenomenon can be deduced from the isospin transport ratio (also known as imbalance ratio) introduced by Rami \textit{et.~al} \cite{Rami_PhysRevLett_84_1120}. 
This technique is based on combining isospin-sensitive observables measured with four systems differing in their initial neutron-to-proton ratio.
Two symmetric neutron-rich (NR) and neutron-deficient (ND) reactions are used as reference values, while the two mixed reactions (M) reach a neutron content in between these two references.

The isospin transport ratio is defined as:

\begin{equation}
R_x = \frac{2 x^{M} - x^{NR} - x^{ND} }{x^{NR} - x^{ND}}
\label{eq_isoratio} 
\end{equation}
where $x$ is an isospin-sensitive observable expected to be univocally related to the $N/Z$ of the systems under study.

In order to follow the evolution of the system towards the $N/Z$ equilibration, the $R_x$ ratio is studied as a function of an ordering parameter directly related to the dissipation.
By construction, $R_{x}=\pm1$ in the limit of fully non-equilibrated conditions (isospin transparency).
Moreover, $R_{x}=0$ represents full isospin equilibration conditions of a mixed reaction if $x$ presents a linear dependence on the isospin, whereas more generally equilibration is signalled by both mixed reactions achieving the same value for the ratio \cite{Rami_PhysRevLett_84_1120}.

In particular, under identical experimental conditions, this ratio is expected to reduce the sensitivity to dynamical (fast emissions, Coulomb) and sequential decay effects while the associated errors remain statistical.
According to the original work of Rami, $R_x$ is expected to bypass any perturbation introducing a linear transformation of $x$. 
The validity of this assumption was recently studied in the framework of the Antisymmetrized Molecular Dynamics (AMD) transport model \cite{ONO2019139} coupled to different statistical decay codes \cite{Camaiani_PhysRevC_102}. 
A conclusion of that work was that the statistical de-excitation of the QP introduces a linear transformation of the $N/Z$ of the projectile for an excitation energy $E^*/A \gtrsim 2$ MeV/nucleon, while a non-linearity is developed at lower excitation energies.

We also address this issue in the next sections with a comparison of the isospin transport ratio obtained from the PLF directly measured in VAMOS and the reconstructed QP with the evaporated neutron estimation.

\subsubsection{Estimation of experimental centrality}\label{subsubsec_expcentrality}

In the present study we will use the total transverse energy of LCP, $E_{t12}$, as a measure of the violence of the collisions in order to estimate their centrality. 
$E_{t12}$ is defined as:
\begin{equation}
E_{t12}=\sum_{i:Z_{i}\leq2}E_{i}\sin^{2}\theta_{i}
\label{eq_et12} 
\end{equation}

where in the sum $i$ runs over the detected products of each event with $Z_{i}\leq2$, laboratory kinetic energy $E_{i}$ and laboratory polar angle $\theta_{i}$. 
This global variable has often been used in previous analyses, as it is particularly well-suited to the performances of the INDRA array for which LCP are detected with a close-to-geometrical (90\%) efficiency \cite{I10-Luk97, I17-Pla99, Ademard_EPJA_50, PhysRevC_104_034609}. 
In the context of the present analysis it is also worth noting that the PLF properties are completely independent of this quantity, avoiding possible trivial bias due to auto-correlation with the event sorting for the study of the isospin transport ratio presented in Section \ref{subsubsec_isoratio}.

\begin{figure}[ht]
\centering
\includegraphics[scale=0.43]{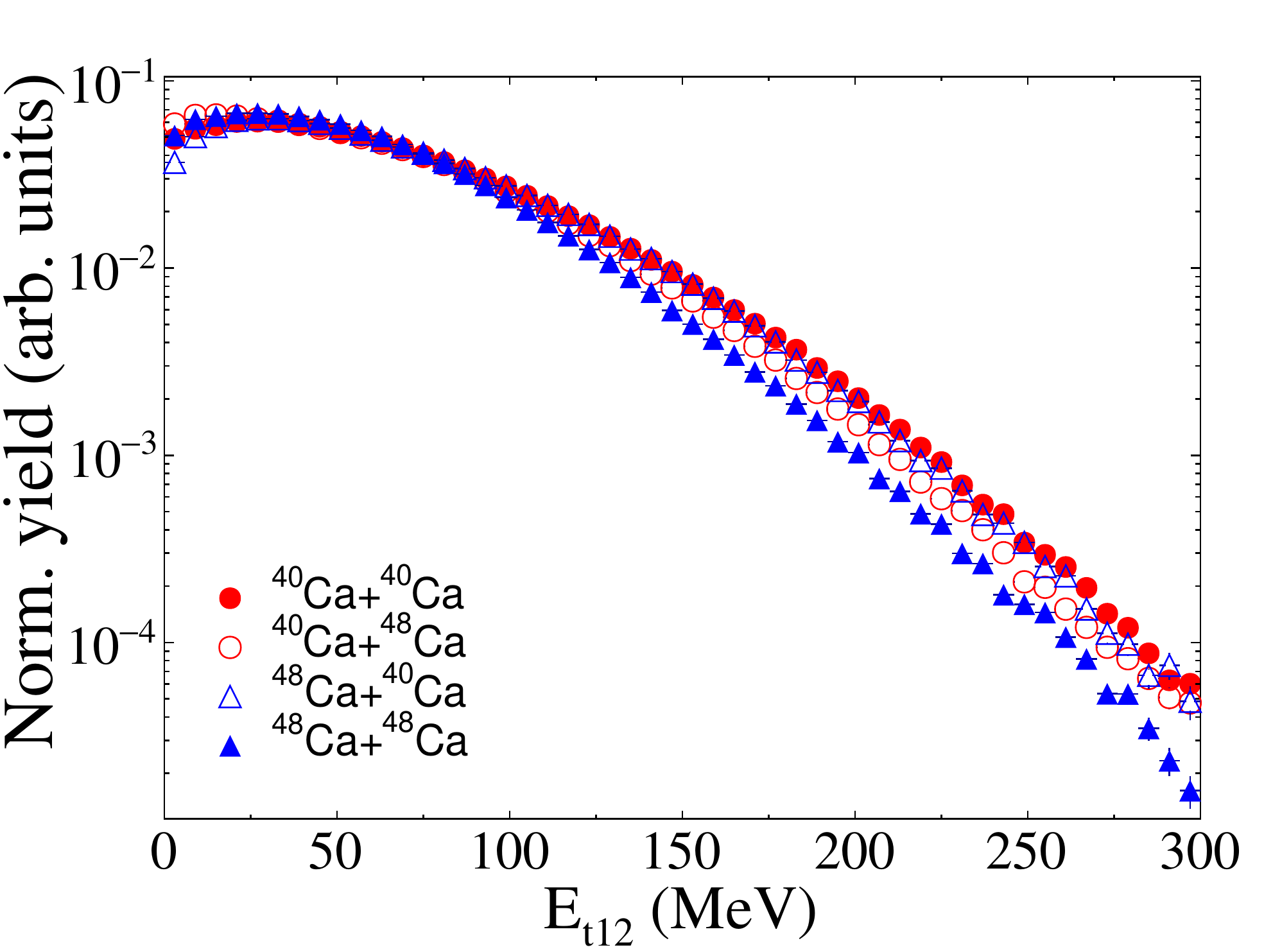}
\caption{(Color online) Total transverse energy distributions of light charged particles (normalized to their integral) for $^{40,48}$Ca$+^{40,48}$Ca collisions at 35 MeV/nucleon.}
\label{fig_Et12}
\end{figure}

The inclusive $E_{t12}$ distributions for the four systems under study are shown in Fig.\ref{fig_Et12}. 
The distributions are presented with a logarithmic $y$-axis scale in order to emphasize the differences in the high-$E_{t12}$ tails, which to a first approximation can be associated with the most
central collisions.
We observe that for the system $^{40}$Ca$+^{40}$Ca higher $E_{t12}$ values are explored in the tail compared to the neutron-rich $^{48}$Ca+$^{48}$Ca one, while the two ``crossed'' systems, $^{40}$Ca+$^{48}$Ca and $^{48}$Ca$+^{40}$Ca, reach intermediate values. 
This means that $E_{t12}$ values for the different systems are not directly comparable, and data for the different systems with similar $E_{t12}$ values cannot be assumed to have similar centrality.

In order to remove the trivial system-dependence of the $E_{t12}$ distributions, $P(E_{t12})$, we will sort data according to the experimental
centrality, $c_{E_{t12}}$, defined as
\begin{equation}
c_{E_{t12}}\equiv\int_{E_{t_{12}}}^{+\infty}P(\tilde{E}_{t12})\:\mathrm{d}\tilde{E}_{t12}\label{eq_cent}
\end{equation}

which is nothing but the complementary cumulative distribution function or tail function of the distribution $P(E_{t12})$. 
By construction, $c_{E_{t12}}$ decreases from 1 to 0 as $E_{t12}$ goes from its minimum ($\sim0$) to its (system-dependent) maximum value, therefore in the following we will associate large ($\sim1$) $c_{E_{t12}}$ values with the most peripheral collisions recorded for each system, while smaller values ($c_{E_{t12}}\rightarrow0$) indicate increasing centrality \textit{i.e.} smaller average impact parameters. 
It should be noted that a further advantage of sorting according to centrality as defined in Eq. \ref{eq_cent} is that $c_{E_{t12}}$ bins of fixed width contain the same number of events or fraction of the total recorded cross-section, and therefore have the same statistical significance, whatever the centrality.

%

\subsubsection{$N/Z$ asymmetry of the projectile}\label{subsubsec_nrichness}

In this section we present the general features of the experimental observable used to compute the isospin transport ratio from Eq.\ref{eq_isoratio}.

The asymmetry $\delta=(N-Z)/A$ of the PLF or the QP was chosen as an isospin sensitive observable as it is proved to be a suitable tool for investigation of the stiffness of the NEoS \cite{Colonna_EPJA50, Ciampi_PRC_106_024603}.
It should be noted that the results and conclusions presented hereafter can also be drawn using the neutron-to-proton ratio $N/Z$. 

\begin{figure}[ht]
\centering
\includegraphics[scale=0.5]{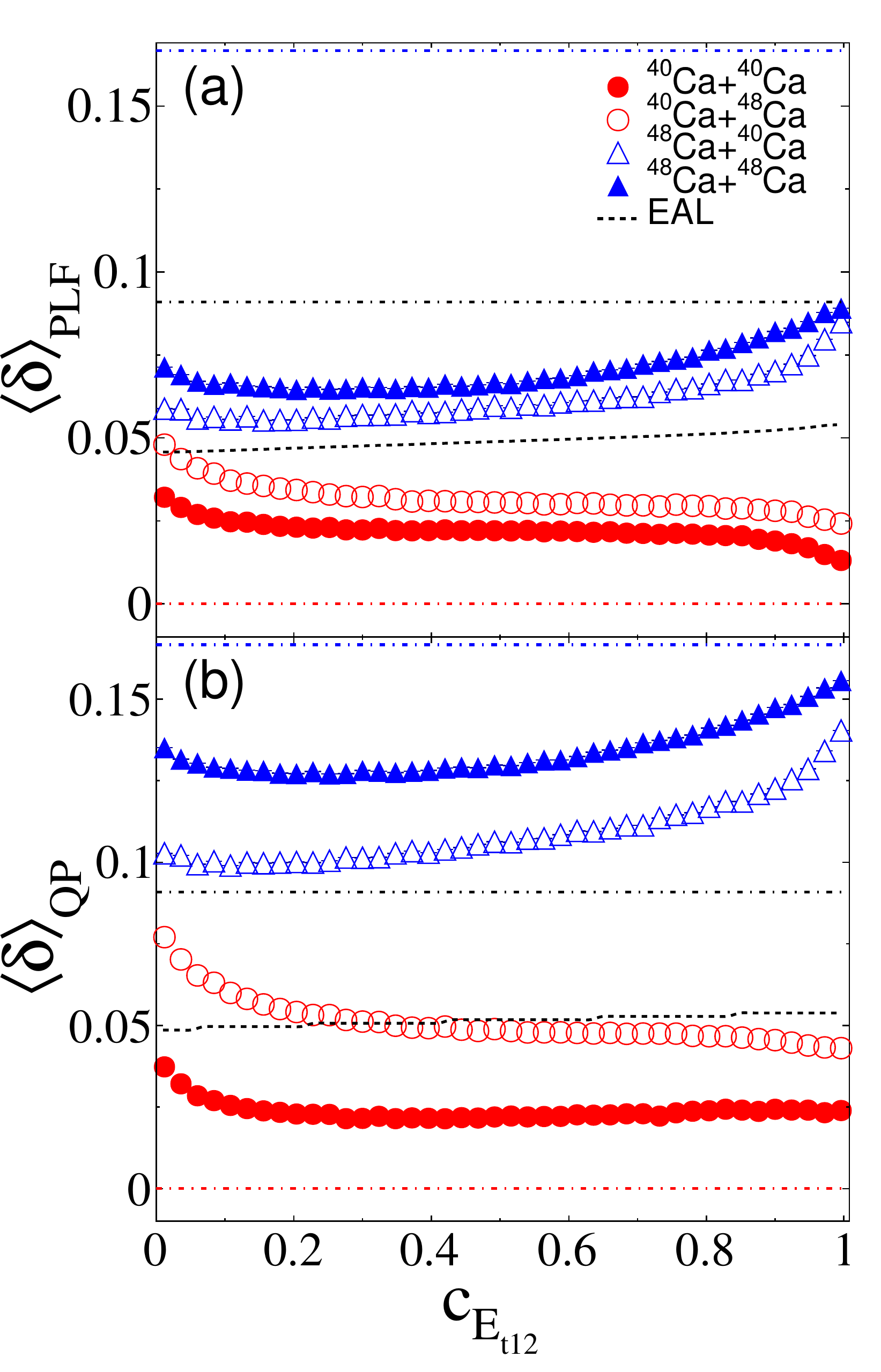}
\caption{(Color online) Distribution of the average asymmetry $\delta$ as a function of the experimental centrality $c_{E_{t12}}$ for (a) the PLF and (b) the reconstructed QP for the four systems under study.
The initial asymmetries of the colliding systems and the EAL extracted from \cite{charity98} (using the average charge of the symmetric systems for each bin of $c_{E_{t12}}$) are represented in dotted-dashed and dashed lines, respectively.}
\label{fig_NmZoA}
\end{figure}

Figure \ref{fig_NmZoA} presents the evolution of the average asymmetry of the PLF and of the reconstructed QP as a function of the experimental centrality $c_{E_{t12}}$.
The PLF distributions, shown in Fig.\ref{fig_NmZoA}(a), exhibit an ordering according to the neutron-richness of the projectile and to a lesser extent of the target, similarly to the charge distributions reported in our previous work \cite{Fable_PRC_106_024605}.
In addition, we observe an increase (respectively a decrease) of the asymmetry with increasing $c_{E_{t12}}$ for the $^{48}$Ca (respectively $^{40}$Ca) projectile reactions.
We also notice that the initial asymmetries ($\delta_{4040}=0$, $\delta_{4048}=\delta_{4840}=0.091$ and $\delta_{4848}=0.167$, represented with dotted-dashed lines) of the colliding nuclei is not reached, even for the less central collisions ($c_{E_{t12}} \simeq 1$).
It is worth noting that for the most peripheral collisions a mean neutron excess of 4 ($\delta=0.091=\delta_{4840}$) is measured for Ca isotopes with $^{48}$Ca projectile reactions, while a mean neutron deficit of 1 is obtained for the $^{40}$Ca projectile.
The average loss of four neutrons for neutron-rich projectile systems can be prevalently associated to neutron evaporation from the excited projectile residue.
An additional contribution from neutron transfer should reflect in an enhanced shift of the isotopic content towards the residue corridor for the asymmetric systems.
Less neutron rich than the $\beta$-stability line, the residue corridor, or evaporation attractor line (EAL, represented here in dashed lines for the symmetric systems and extracted from \cite{charity98}), is a region of the nuclear chart where proton and neutron emissions have equal probability.
It is reached when enough excitation energy feeds the decay, rather than other more violent decay mechanisms.
For the $^{48}$Ca projectiles, this effect even counterbalances the fact that the asymmetric system should correspond to less excitation energy available for the secondary decay (due to the smaller target) so that, in absence of diffusion and relying on a simple sequential-decay picture, the residues produced with the $^{48}$Ca+$^{48}$Ca system should approach the EAL more closely than the $^{48}$Ca+$^{40}$Ca system.
The opposite is observed in the present case, likely due to isospin diffusion contributions.
For the $^{40}$Ca projectiles, the two effects should on the other hand act in phase.
It is interesting to mention that the opposite effect was observed at relativistic energies in \cite{Napolitani_PRC_76_064609, Henzlova_PRC_78_044616}, where a prevalent contribution from multifragmentation was taking excitation away from the secondary decay process, preventing the system from reaching the EAL.
A quantitative description of the distributions of Fig. \ref{fig_NmZoA}(a) should also involve the emission of light clusters, in particular for the $^{40}$Ca system and the side related to larger centrality, as more excitation energy would feed cluster emission.

Concerning the reconstructed QP asymmetry presented in Fig.\ref{fig_NmZoA}(b), the same trend and hierarchy than for the PLF distributions are observed.
As expected, the reconstructed $\delta$ are further away of the EAL than in the case of the PLF with higher gaps between the distributions.
We remark that the $^{48}$Ca projectile reactions present systematically higher values than the initial asymmetry of the asymmetric reactions ($\delta_{4048}=\delta_{4840}=0.091$) with lower values than the neutron rich system ($\delta_{4848}=0.167$).
For the most peripheral collisions a mean neutron excess of 6 ($\delta=0.130$) is measured for Ca isotopes with $^{48}$Ca projectile reactions, while a mean neutron deficit of $0.5$ is obtained for the $^{40}$Ca projectile.
Finally, it is worth noting that a tendency to reach the initial asymmetry with decreasing $c_{E_{t12}}$ (more central collisions) is observed for the asymmetric systems.

%

\subsubsection{Isospin transport ratio}\label{subsubsec_isoratio}

\begin{figure}[ht]
\centering
\includegraphics[scale=0.42]{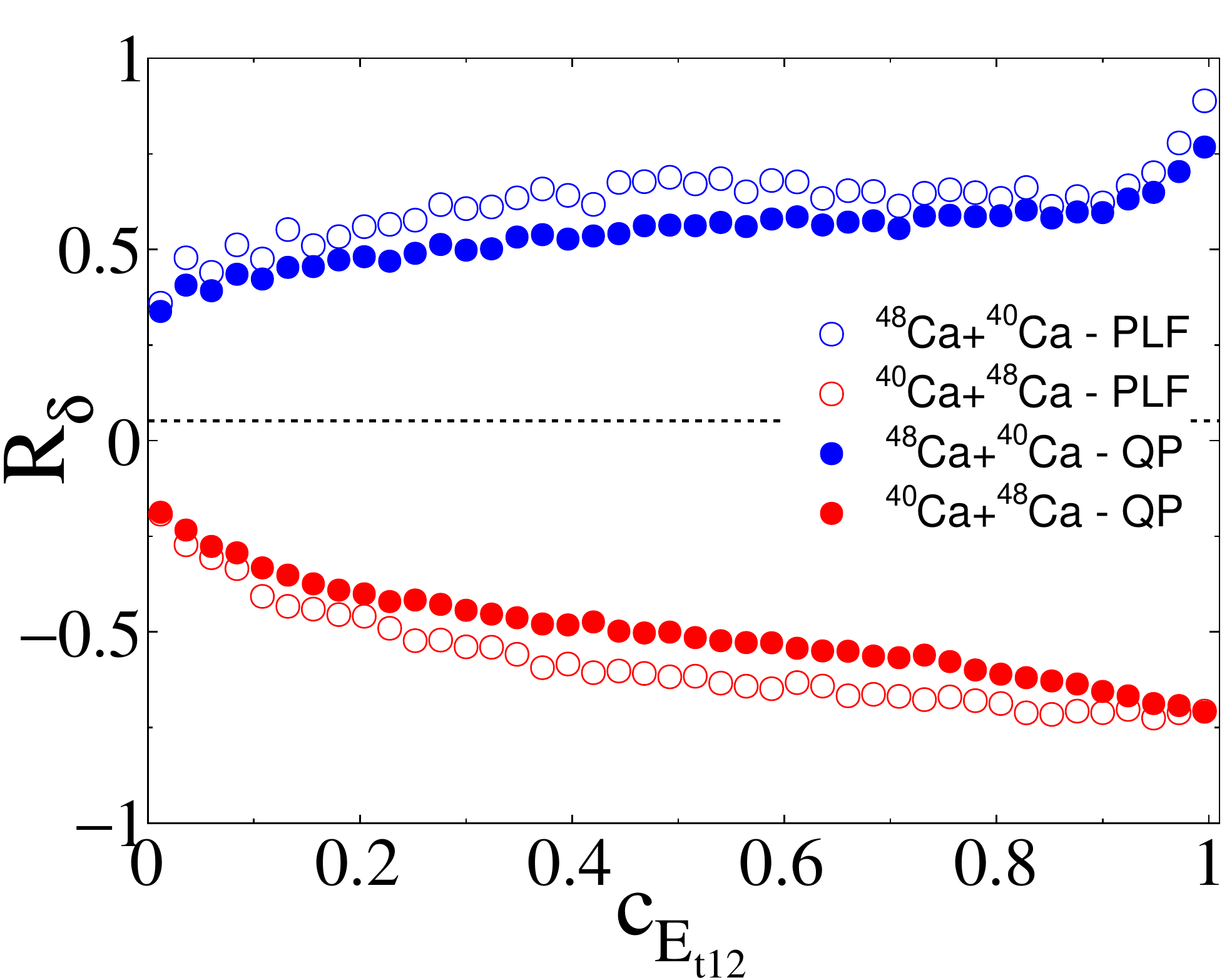}
\caption{(Color online) Isospin transport ratios computed from the asymmetry $\delta$ of the QP remnant (open symbols) and the reconstructed QP (full symbols) as a function of the experimental centrality $c_{E_{t12}}$.
The dashed line correspond to the value ($R_{\delta}=0.04$) at halfway between the $^{40}$Ca+$^{48}$Ca and $^{48}$Ca+$^{40}$Ca experimental points.}
\label{fig_RNmZoA}
\end{figure}

For each bin of experimental centrality $c_{E_{t12}}$, the isospin transport ratio $R_{\delta}$ was computed from Eq.\ref{eq_isoratio} using the average asymmetry $\langle \delta \rangle = \langle (N-Z)/A \rangle$ as the isospin-sensitive observable, with $^{40}$Ca+$^{48}$Ca or $^{48}$Ca+$^{40}$Ca as mixed systems (M), $^{48}$Ca+$^{48}$Ca as neutron-rich system (NR) and $^{40}$Ca+$^{40}$Ca as neutron-deficient system (ND).
The results are presented in Fig.\ref{fig_RNmZoA} for the PLF (open symbols) and the reconstructed QP (full symbols).
For both, an evolution of the isospin transport ratio is observed, with decreasing values from about $R_{\delta} = \pm 0.75$ to about $R_{\delta} = \pm 0.3$ when moving from the most peripheral collisions (high $c_{E_{t12}}$ values) to the most central collisions (low $c_{E_{t12}}$ values).
Moreover the value at halfway between the mixed reactions points (represented in dashed line) is very small ($R_{\delta}=0.04 \pm 0.05$ for the reconstructed QP and $R_{\delta}=0.04 \pm 0.07$ for the PLF), demonstrating the linear dependence of $\delta$.
The data thus suggest a continuous evolution towards an $N/Z$ equilibration condition, while a full equilibration condition is not reached even in the most central collisions. 
It is worth noting that indirect indications, obtained from mirror nuclei yield ratios, of such an evolution of the isospin transport ratio with centrality were already reported in \cite{Sun_PhysRevC_82_051603}.
However, the results reported in the present article are a direct estimate of the $N/Z$ of both the PLF and the QP performed with the coupled INDRA-VAMOS device.

It should be also highlighted that the reconstructed QP presents a smoother evolution, with less disparities and a slower evolution to the full equilibration, compared to the case of the PLF.
As demonstrated in \cite{PhysRevC_98_044602,Ciampi_PRC_106_024603}, for a given sorting parameter, the experimental isospin transport ratio can present variations in its absolute values depending on the isospin-sensitive probe, while the trend towards isospin equilibration was nonetheless observed.

\begin{figure*}[ht]
\centering
\includegraphics[scale=0.87]{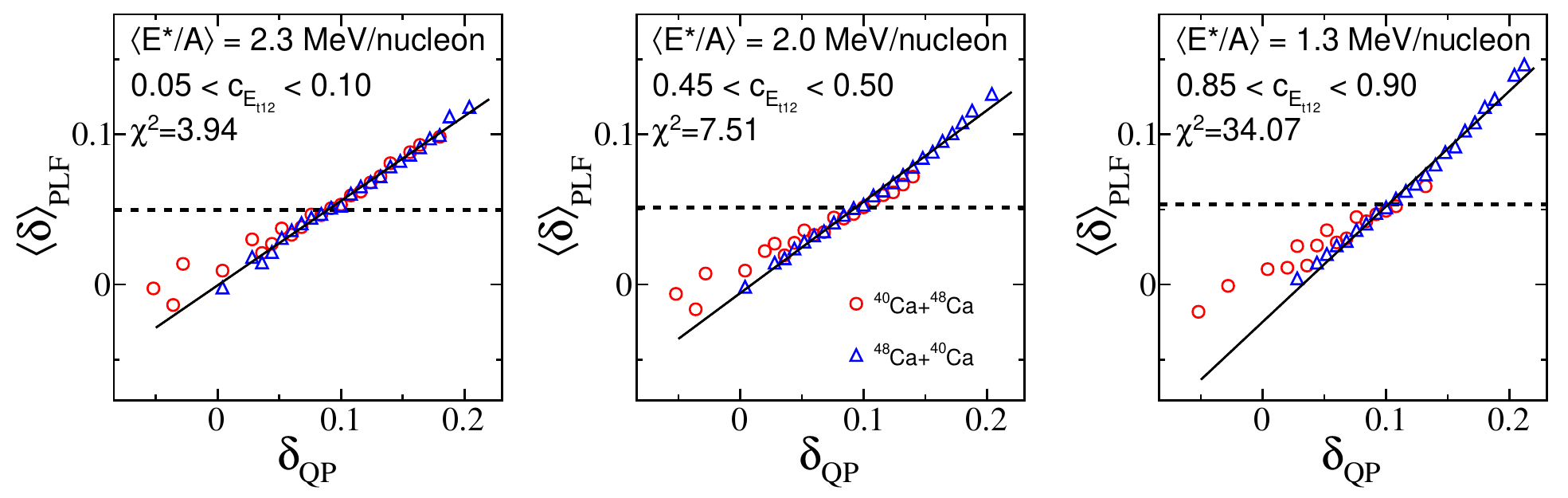}
\caption{(Color online) Correlations between the asymmetry obtained from the QP remnant measured with VAMOS and the reconstructed QP for the mixed systems and different domains of $c_{E_{t12}}$, for the most abundant nuclei.
The continuous and dashed lines represents the linear fits and values predicted by the EAL for the $^{48}$Ca$+^{40}$Ca system, respectively.}
\label{fig_NmZoA_linearity}
\end{figure*}

In order to study the effect of the QP reconstruction on the experimental isospin transport ratio, we present in Fig.\ref{fig_NmZoA_linearity} the average PLF asymmetry, $\langle\delta\rangle_{PLF}$, as a function of the QP asymmetry, $\delta_{QP}$, for three ranges of $c_{E_{t12}}$ values, namely $0.05 < c_{E_{t12}} < 0.1$ (central collisions, left panel), $0.45 < c_{E_{t12}} < 0.5$ (mid-central collisions, central panel) and $0.85 < c_{E_{t12}} < 0.9$ (peripheral collisions, right panel.
In particular, $\langle\delta\rangle_{PLF}$ was computed for each bin of $\delta_{QP}$ for the most abundant nuclei (those which contribute the most to the computation of the isospin transport ratio).
The corresponding average excitation energy for each domain of $c_{E_{t12}}$ is also reported on the figure.

Concerning the $^{48}$Ca$+^{40}$Ca system (blue open triangles), we observe a behaviour similar to the one reported in \cite{Camaiani_PhysRevC_102}: the higher the average excitation energy of the QP, the closer the final QP remnant is to the EAL (dashed horizontal line).
Futhermore, we observe that the reduced chi-squared $\chi^2$ obtained from a linear fit to the $^{48}$Ca$+^{40}$Ca data (continuous line) increases with decreasing average excitation energy, showing more particularly that the correlation is not linear for the less dissipative collisions where $\langle E^{*}/A\rangle=1.3$ MeV/nucleon.

Concerning the $^{40}$Ca$+^{48}$Ca system (red open dots), the interpretation is not straight-forward as a clear linearity of the correlation between $\langle\delta\rangle_{PLF}$ and $\delta_{QP}$ is not verified. 
We observe nonetheless a general evolution of the asymmetry to higher values for both the QP and the PLF with increasing average excitation energy, so that the values increasingly cross the EAL.
 
According to the aforementioned observations, we believe the reconstructed QP asymmetry to be more relevant in order to directly estimate the transport coefficients \cite{PhysRevC_72_064620}.

%

\subsection{Isospin migration}\label{subsec_IsospinMigration}

As previously discussed, a possible indication for the isospin migration is a neutron-enrichment in the neck region induced by a local density gradient in the portion of nuclear matter formed during the collision.

To study this phenomenon experimentally, we focus on the average neutron-to-proton ratio $\langle N/Z \rangle_{CP}$ of the clusters isotopically identified with INDRA and defined as:

\begin{equation}\label{eq_sumN_sumZ_CP}
\langle N / Z \rangle_{CP} = \frac{\sum_{\nu} M_{\nu}\frac{N_{\nu}}{Z_{\nu}}}{\sum_{\nu} M_{\nu}}
\end{equation}
where $M_{\nu}$, $N_{\nu}$ and $Z_{\nu}$ are respectively the multiplicity, neutron number and proton number of each cluster $\nu$, where $\nu$ corresponds to $^{2,3}$H, $^{3,4,6}$He, $^{6,7,8,9}$Li or $^{7,9,10}$Be isotopes (free protons and - undetected - neutrons are excluded).  
Thus, by construction, a neutron-richness would be indicated by $\langle N / Z \rangle_{CP} > 1$.

In particular, we studied the clusters emitted (i) forward with respect to the PLF and (ii) in an angular region centered around the mid-rapidity of the reaction.
The forward domain of the PLF, corresponding to clusters with a positive parallel velocity in the PLF reference frame ($Vz^{\nu}_{PLF}>0$), is expected to be mostly populated by the decay emissions of the QP, minimizing contamination by the QT and neck (dynamical) emissions.
The mid-rapidity domain, corresponding to an angular region centered around mid-rapidity in the reaction center of mass frame ($85^{\circ} < \theta^{\nu}_{CM} < 95^{\circ}$) is expected to be dominated by neck emissions.

The reduced velocity of the PLF, defined as $V_{red}^{PLF} = Vz^{PLF}/V^{proj}$ were $Vz^{PLF}$ and $V^{proj}$ are respectively the parallel velocity of the PLF and the projectile velocity in the laboratory, was chosen as a surrogate for the measure of the degree of dissipation of the collisions.
It also presents the advantage of being independent of the $\langle N / Z \rangle_{CP}$ computed with the LCP identified with INDRA, avoiding trivial bias due to autocorrelation with the event sorting.
It should be noted that the results and conclusions presented hereafter can also be drawn using various $N/Z$ ratios for complex particles described in the literature, such as the one defined in \cite{I71-Gal09, Ademard_EPJA_50}.

\begin{figure*}[ht]
\centering
\includegraphics[scale=0.88]{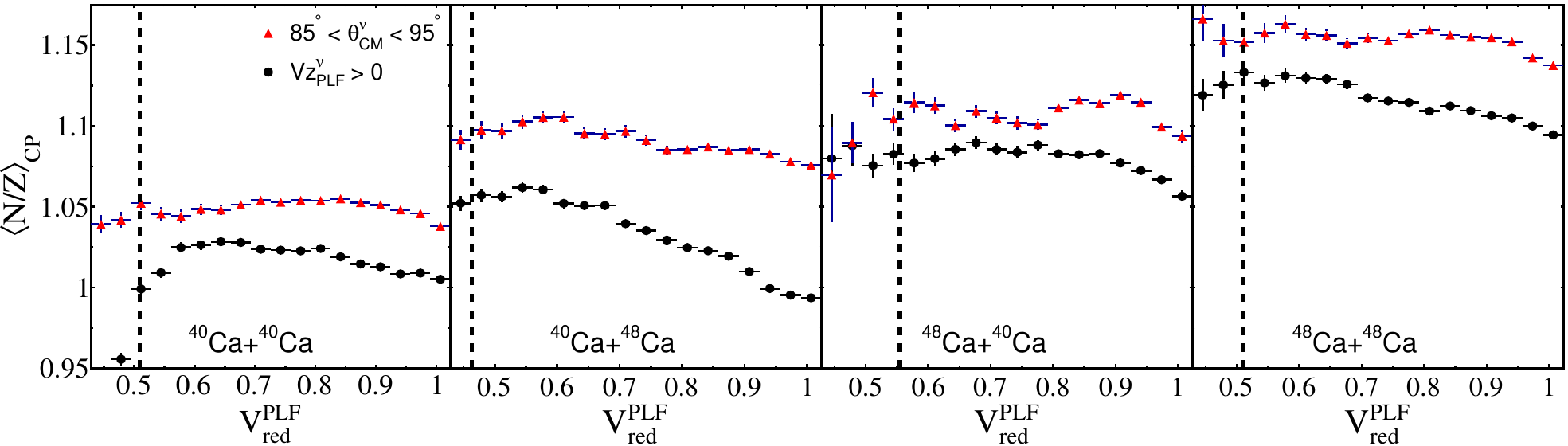}
\caption{(Color online) 
Average $\langle N / Z \rangle_{CP}$ (see Eq.\ref{eq_sumN_sumZ_CP}) for each bin of PLF reduced velocity, for the clusters emitted forward with respect to the PLF (full circles) and in the angular region centered around the mid-rapidity (full triangles). Center of mass reduced velocity is represented with a dotted line.}
\label{fig_SumNSumZ}
\end{figure*}

The results are presented in Fig.\ref{fig_SumNSumZ}, where the average $\langle N / Z \rangle_{CP}$ is plotted as a function of $V_{red}^{PLF}$.

Concerning the forward emission region, we remark that the $\langle N/Z \rangle_{CP}$ tends to follow the hierarchy of the asymmetry $\delta$ observed in Fig.\ref{fig_NmZoA}. 
This suggests that the measured $N/Z$ ratio for complex particles can also be exploited as an alternative probe to evaluate the isospin transport ratio in peripheral collisions with the INDRA-VAMOS coupling.

Concerning emissions at mid-rapidity, we observe higher $\langle N/Z \rangle_{CP}$ values compared to the foward domain, independently of the dissipation. 
This suggests a systematic neutron enrichment of the mid-rapidity region, which is dominated by the density drop at mid-rapidity rather than by the isotopic content of the colliding partners, in particular when they are symmetric.
Such observation can be interpreted as a consequence of isospin migration. 
Furthermore, free protons, most of which originate from prompt out-of-equilibrium emissions, are in fact excluded in the computation of $\langle N/Z \rangle_{CP}$.
Finally, an overall smoother increase of $\langle N/Z \rangle_{CP}$ is observed with increasing dissipation for the mid-rapidity region compared to the forward region.

\begin{figure}[ht]
\centering
\includegraphics[scale=0.265]{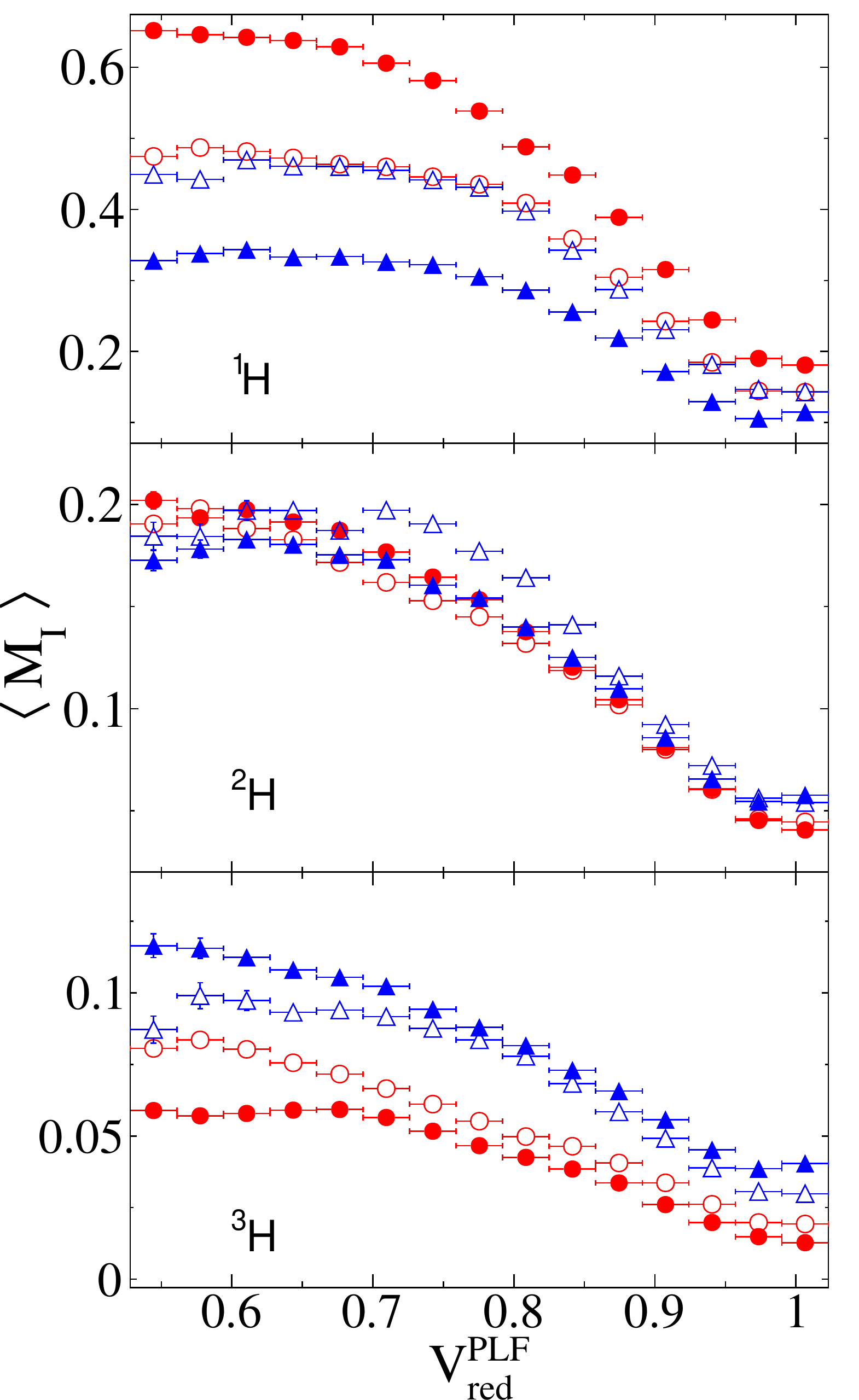}
\includegraphics[scale=0.265]{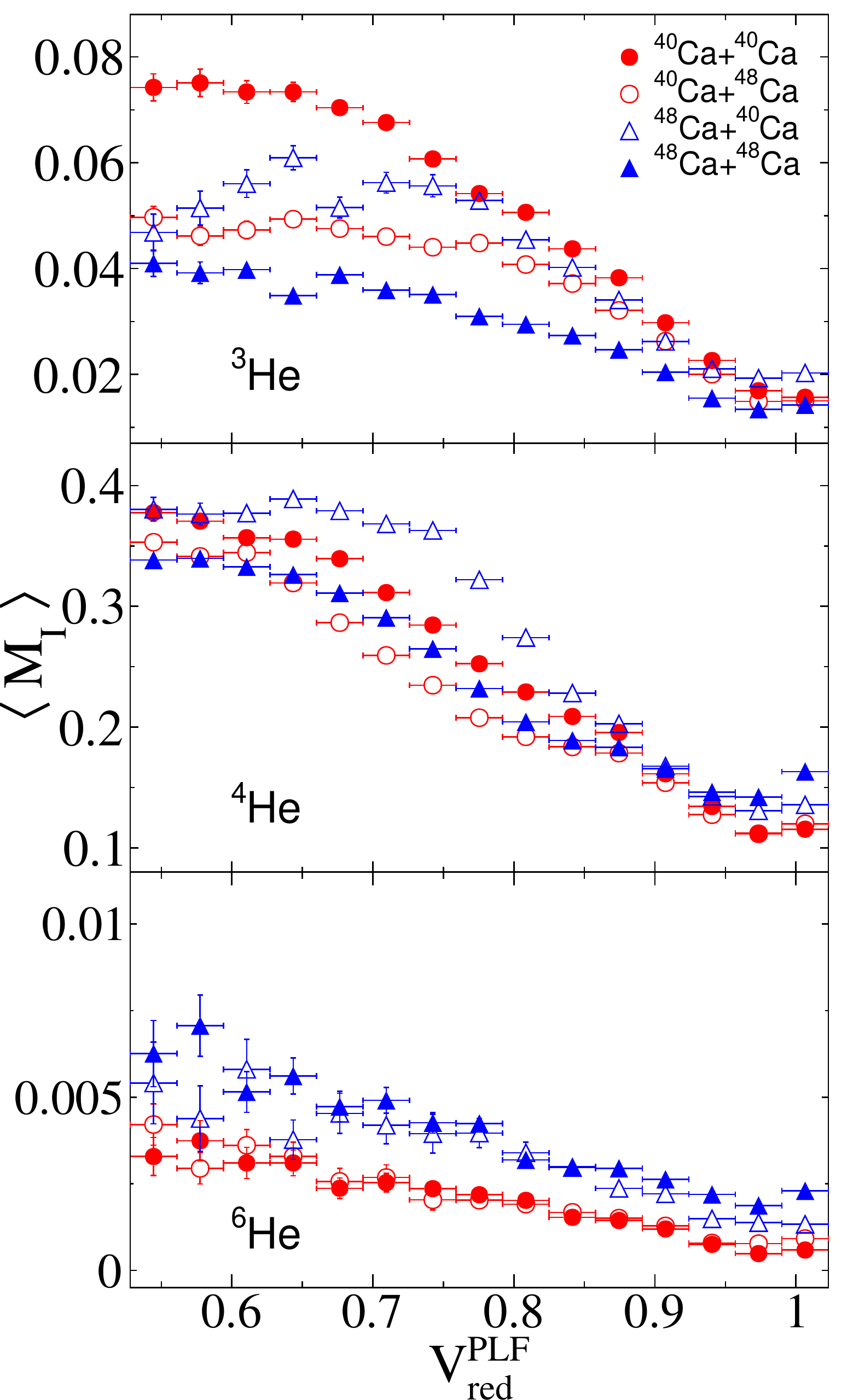}
\caption{(Color online) Average LCP multiplicities detected in INDRA as a function of the QP remnant reduced velocity for the mid-rapidity domain (see text).}
\label{fig_MZ1_Z2}
\end{figure}

In order to further investigate the individual contributions to the mid-rapidity region, we present in Fig.\ref{fig_MZ1_Z2} the average multiplicities of the LCP that contribute the most to the computation of $\langle N/Z \rangle_{CP}$, as a function of the reduced velocity of the PLF.
For the sake of concision, the proton multiplicities are also represented here but we remind that they are not included in Eq.\ref{eq_sumN_sumZ_CP}. 

As expected, the multiplicities increase with decreasing $V_{red}^{PLF}$, reflecting an increase of cluster production with increasing dissipation as more excitation energy is available.
We observe that for the protons and neutron-poor $^{3}$He particles, the multiplicities mostly reflect the total (projectile and target) neutron content of the initial colliding system, leading to close distributions for the mixed $^{40}$Ca$+^{48}$Ca and $^{48}$Ca$+^{40}$Ca systems in the case of more dissipative collisions.	
We would like to highlight that this pattern is very different from the one obtained for the forward domain (see Fig.8 from \cite{Fable_PRC_106_024605}) where a hierarchy related, first, to the neutron richness of the projectile and, to a lesser extent, of the target is observed.
Concerning the neutron-rich tritons and $^{6}$He particles, we notice that the distributions mostly depend on the projectile neutron-richness. 
Finally, in the case of the deuterons and $^{4}$He particles, having the same $N/Z$, we observe that the multiplicities depend much less on the system, similarly to the forward domain, with a noticeable difference for the $^{48}$Ca$+^{40}$Ca system.
Surprisingly, we also remark that the absolute values are close for the symmetric systems independently of the centrality.

As a conclusion, the trends observed in Fig.\ref{fig_MZ1_Z2} suggest that the behaviour of $\langle N/Z \rangle_{CP}$, thus the neutron-enrichment of the mid-rapidity domain, is mainly driven by the $A=3$ LCP isobars.
Indeed, contrary to what we observe for the forward velocity domain, the $^{3}$He and $^{3}$H multiplicities mainly depend on the neutron-richness of the total system or the projectile, respectively.
Thus, for the symmetric system, the behaviour is straightforward and independent of the dissipation as the neutron-richness of the total system and the projectile act in phase, leading to higher values of $\langle N/Z \rangle_{CP}$ for the $^{48}$Ca$+^{48}$Ca reaction.
For the asymmetric systems, the effect is more complex as the $^{3}$He multiplicities tend to converge with increasing centrality, leading to values of $\langle N/Z \rangle_{CP} \simeq 1.1$ for both systems.

The above observations suggest that the experimental isotopic ratios reflect a neutron enrichment in the mid-rapidity velocity region.
This can be interpreted as consequence of isospin migration, confirming that the density at mid-rapidity is lower than the saturation density $\rho_0$, leading the neutrons to be attracted towards this zone.
Also, similarly to the study given in \cite{BOUGAULT_PhysRevC_97_024612} for heavier partners $^{136,124}$Xe$+^{124,112}$Sn reactions, we believe that comparisons of the triton and $^{3,6}$He particles production with filtered transport model calculations, more specifically for peripheral collisions, could lead to a better understanding of isospin-dependence of the NEoS.  

\section{Conclusion}\label{sec_conclusion}
In this work we investigated the isospin transport phenomena from $^{40,48}$Ca+$^{40,48}$Ca reactions at 35 MeV/nucleon measured with the INDRA-VAMOS coupling.
This apparatus allows to reconstruct the hot quasiprojectile source from the remnant (PLF) directly identified with the VAMOS high acceptance spectrometer and carefully selected light-charged-particles (LCPs) measured in coincidence with INDRA, on an event-by-event basis.

The isospin diffusion phenomena was studied from the isospin transport ratio $R_{\delta}$, using the asymmetry $\delta=(N-Z)/A$ as the isospin sensitive observable and the experimental centrality calculated from the total transverse energy of the LCPs with $Z \leqslant 2$, $c_{E_{t12}}$, as a sorting parameter to appraise the violence of the collision.
Both observables present the interest of being uncorrelated, avoiding any bias in $\delta$ from a bin selection in $c_{E_{t12}}$.
For both QP and PLF, a deacrease of the isospin transport ratio is measured with decreasing $c_{E_{t12}}$, from about $R_{\delta} = \pm 0.75$ for the most peripheral collisions to about $R_{\delta} = \pm 0.3$ for the most central.
Thus, the overall isospin transport ratios present, for both the PLF and the reconstructed QP, a clear evolution towards isospin equilibration with increasing dissipation of the collision (decreasing $c_{E_{t12}}$), while a full $N/Z$ equilibration condition is not reached.
Furthermore, a smoother evolution, with less discrepancies between the two mixed systems, is observed in the case of the reconstructed QP.

The effect of the reconstruction of the QP on the experimental isospin transport ratio was also studied from the correlations between the QP asymmetry $\delta_{QP}$ and the average PLF asymmetry $\delta_{PLF}$.
Concerning the $^{48}$Ca$+^{40}$Ca system, we noticed that the correlation is not linear for the less dissipative collisions where $\langle E^{*}/A \rangle =1.3$ MeV/nucleon.
We also observed that the higher the average excitation energy of the QP, the closer the final QP remnant is to the EAL. 
Concerning the $^{40}$Ca$+^{48}$Ca system, we observe an evolution of the asymmetry to higher values for both the QP and the PLF with increasing average excitation energy.

The isospin migration was studied from the average neutron-to-proton ratio $\langle N/Z \rangle_{CP}$ of $^{2,3}$H, $^{3,4,6}$He, $^{6,7,8,9}$Li or $^{7,9,10}$Be clusters isotopically identified with INDRA (free protons excluded). 
We focused more specifically on the clusters emitted in two distinct angular domains: the one forward-emitted with respect to the PLF (supposed to correspond to the decay emissions of the QP) and the one in an angular region centered around mid-rapidity in the reaction center of mass frame (supposed to characterize the neck emissions).
The reduced velocity $V_{red}^{PLF}$ of the PLF measured with VAMOS, measured independently of the $\langle N/Z \rangle_{CP}$ observable, was used as a sorting parameter to appraise the violence of the collision.
A systematic neutron-enrichment ($\langle N/Z \rangle_{CP}>1$) of the mid-rapidity domain compared to the forward domain was observed for all systems, independently of the sorting parameter.
The aforementioned observation can be interpreted as a consequence of isospin migration for the symmetric systems as they present no isospin gradient between the projectile and the target.
Furthermore, the study of the $Z=1$ and $Z=2$ LCPs multiplicity distributions allowed to highlight that the behaviour of $\langle N/Z \rangle_{CP}$ is mainly driven by the $A=3$ isobars for the mid-rapidity domain, while $^{3}$He and $^{3}$H multiplicities mainly depend on the neutron-richness of the total (projectile and target) system or the projectile, respectively.
Thus, if the behaviour of the symmetric systems is straightforward and (mostly) independent of the centrality, the effect is more complex for the asymmetric ones as the $^{3}$He multiplicities tend to converge with increasing centrality, leading to values of $\langle N/Z \rangle_{CP} \simeq 1.1$.
Finally, it must also be noted that the $\langle N/Z \rangle_{CP}$ of the forward domain tends to follow the same hierarchy as the asymmetry $\delta$, suggesting that such observable can also be exploited as an alternative probe to evaluate the isospin transport ratio in peripheral collisions.

The results presented in this work demonstrate the relevance of the data collected with the INDRA-VAMOS coupling and its high potential to provide further information and constraints on the study of the symmetry energy term in the NEoS for peripheral HIC collisions. 
Extensive comparisons of this dataset with various dynamical and statistical model calculations are foreseen, in order to better understand and constrain isospin transport phenomena and more generally the dynamical processes at stake.
In particular, it would be interesting to track the density evolution for such collisions within an accurate mean-field description in the presence of dynamical instabilities.

\begin{acknowledgments}
The authors would like to thank:
The technical staff of GANIL for their continued support for performing the experiments; 
A. Navin for his constant support; M. Rejmund for setting up the VAMOS spectrometer;
B. Lommel and the Target Laboratory of the GSI Helmholtzzentrum for providing the $^{48}$Ca targets;
The Target Laboratory of Legnaro for providing the $^{40}$Ca targets;
W. Catford for providing TIARA electronics used for the CsI wall in the VAMOS focal plane; 
A. Lemasson and B. Jacquot for their invaluable help with VAMOS trajectory reconstructions;
The CNRS/IN2P3 Computing Center (Lyon - France) for providing data-processing resources needed for this work.
Q. F. gratefully acknowledges the support from CNRS-IN2P3.
\end{acknowledgments}


\bibliographystyle{apsrev}
\addcontentsline{toc}{section}{\refname}
\bibliography{extracted.bib}

\end{document}